\documentclass[aps,prl,twocolumn,superscriptaddress,showpacs,floats,floatfix,preprintnumbers,amsmath,amssymb]{revtex4}
\usepackage{graphicx}
\usepackage{dcolumn}
\usepackage{bm}

\begin{document}

\preprint{(to be submitted to Phys. Rev. Lett.)}

\title{The energy landscape of fullerene materials: a comparison between boron, boron-nitride and carbon }

\author{Sandip De}
\affiliation{Department of Physics, Universit\"{a}t Basel, Klingelbergstr. 82, 4056 Basel, Switzerland}
\author{Alexander Willand}
\affiliation{Department of Physics, Universit\"{a}t Basel, Klingelbergstr. 82, 4056 Basel, Switzerland}
\author{Maximilian  Amsler}
\affiliation{Department of Physics, Universit\"{a}t Basel, Klingelbergstr. 82, 4056 Basel, Switzerland}
\author{ Pascal Pochet}
\affiliation{Laboratoire de simulation atomistique (L\_Sim), SP2M, UMR-E CEA / UJF-Grenoble 1, INAC, Grenoble, F-38054, France}
\author{Luigi Genovese}
\affiliation{Laboratoire de simulation atomistique (L\_Sim), SP2M, UMR-E CEA / UJF-Grenoble 1, INAC, Grenoble, F-38054, France}
\author{Stefan Goedecker}
\affiliation{Department of Physics, Universit\"{a}t Basel, Klingelbergstr. 82, 4056 Basel, Switzerland}

\date{\today}

\begin{abstract}
Using the minima hopping global geometry optimization method on the density functional potential energy surface we show that the 
energy landscape of boron clusters is glass like. 
Larger boron clusters have many structures which are lower in energy than the cages. 
This is in contrast to carbon and boron nitride systems which can be clearly identified as structure seekers.  
The differences in the potential energy landscape explain why carbon and boron nitride systems 
are found in nature whereas pure boron fullerenes have not been found. We thus present a methodology which can make predictions 
on the feasibility of the synthesis of new nano structures.

\end{abstract}


\maketitle

The experimental synthesis of fullerenes is a very difficult task. The carbon fullerene structures were therefore theoretically predicted~\cite{predict} long  before they could be produced in the lab~\cite{c60discovery}. 
Many more hollow and enhodedrally doped fullerene structures made 
out of elements different from carbon have also been proposed since then theoretically~\cite{list} in searches of other possible building blocks for nano-sciences. 
It is however surprising that since the experimental discovery of the carbon fullerenes some 25 years 
ago no other fullerenes have been synthesized. So the question is whether experimentalists have just not yet found a way to synthesize these theoretically predicted fullerenes, or whether they do not exist at all in nature.
We have recently shown~\cite{si20} that all the theoretically proposed endohedral $Si_{20}$ fullerenes are meta-stable and can thus most likely not be found in nature.
In this letter we investigate in detail boron clusters. 
Following the $B_{80}$ fullerene structure proposed by Szwacki et al.~\cite{Szwacki} various other fullerene~\cite{Bfull} and stuffed fullerene 
structures ~\cite{sBfull} were proposed. Subsequently it was however shown for $B_{80}$ that there exist non-fullerene structures~\cite{chinese} which are lower in energy. 
We will contrast the characteristics of the potential energy landscape (PES) of these boron clusters with those of systems found in nature, 
namely carbon and boron nitride fullerenes and find that there are important differences. 

To explore the energy landscape of the boron, carbon and boron nitride clusters we do global geometry optimizations on the density 
functional potential energy surface with the minima hopping algorithm~\cite{minhop}.  This algorithm
can render the global minimum configuration as well as many other low energy meta-stable structures. 
All the density functional calculations are done with the BigDFT electronic structure code~\cite{bigdft} which uses a systematic wavelet basis 
together with pseudopotentials~\cite{gth} and the standard LDA~\cite{gth} and PBE~\cite{pbe} exchange correlation functionals. 

\begin{figure} [!h]
\includegraphics[width=0.35\textwidth]{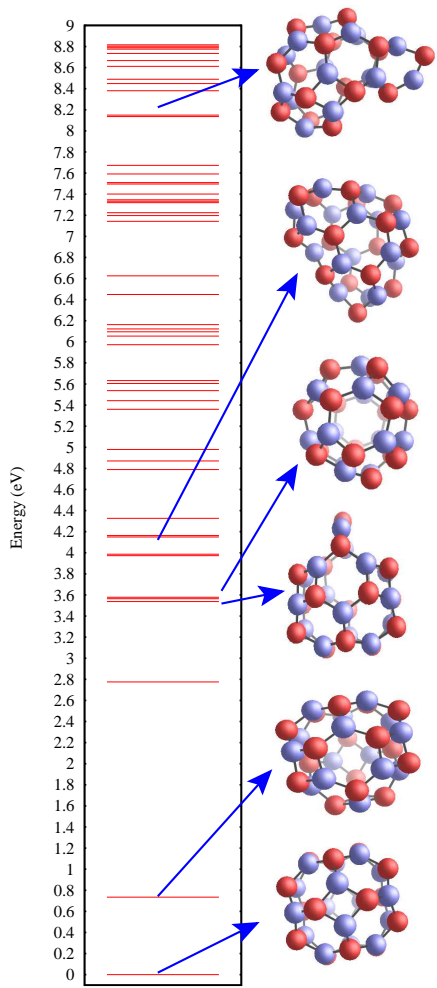}
\caption{\label{B16N16} The configurational energy spectrum of $B_{16}N_{16}$. Boron atoms are shown in blue  and nitrogen atoms in red.
The higher energy cage structures can be described as a `basket' with a `handle' made out of a chain of 4 atoms (two of each type). }
\end{figure}

We start out by analyzing the $B_{16}N_{16}$ cluster which was found to be short lived in experiments~\cite{goldberg}. In this system structural 
rigidity is imposed by a strong preference for sp2 hybridization~\cite{trend} 
as well as by the requirement that bonds are only formed between atoms of different type.
This leads to a small configurational density of states. 
As shown in Fig.~\ref{B16N16} there exists  a fairly large energy interval in which only cage like structures exist.
Hence there is a strong driving force towards the ground state cage structure and minima hopping can find it rapidly. 
This driving force also allows the formation of $B_{16}N_{16}$ in nature. 

\begin{figure}
\includegraphics[width=0.45\textwidth]{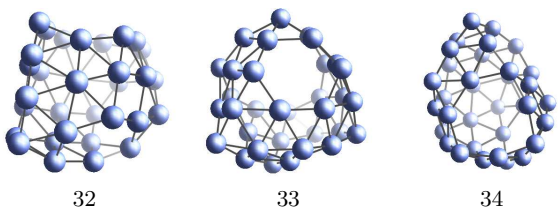}
\caption{\label{B_medium} Global minima of $B_{32}$,$B_{33}$ and $B_{34}$ }
\end{figure}

Using the same methods we went on to study medium size boron clusters with 32-36 atoms. 
In this size range the clusters show a strong tendency to form cages and all the numerous low energy structures we found are 
cage like.  This is agreement with a recent study~\cite{medium} where the ground state was found to be cage like. 
Three representative ground state structures are shown in Fig.~\ref{B_medium}.
These medium size  clusters contain well known structural motifs~\cite{boustani} namely  empty and filled hexagons as well as 
empty and filled pentagons. But in addition they contain numerous other structural motifs such as single atoms connecting filled 
hexagons or rings containing more  than 6 atoms. The inclusion of these other structural motifs does not rise the 
energy significantly and the first meta-stable structure is typically only 0.1 eV higher in energy than the global minimum. 
For $B_{32}$ we found for instance some 100 cage like isomers in an energy interval of only 
1 eV above the global minimum and even more isomers presumably exist in this interval. The number of nearest neighbors in these 
structures varies from 4 to 6 and the bond angles vary from 90 degrees for some 4 fold coordinated corner atoms to 60 degrees for 
6 fold coordinated atoms in the center of a planar hexagon. This is in contrast to the structural rigidity 
imposed by the sp2 hybridization on all the carbon fullerene structures we have generated. Even though one can find in our 60 atom carbon structures 
rectangles and heptagons in addition to hexagons and pentagons, all the atoms have, without any exception, 3 nearest neighbors 
in structures that are less than 20 eV above the ground state. As a consequence 
we expect the configurational density of states to be much smaller for carbon cages than for boron clusters. 
This is indeed the case as will be shown in more detail later. 

Next we did global geometry optimization runs for the $B_{80}$ cluster. 
A first run started from the Szwacki fullerene, which consists of the $C_{60}$ fullerene with 
20 additional atoms filled into the hexagons.  It thus consists of 20 filled hexagons and 12 empty pentagons. 
The insertion of the 20 atoms can be viewed as some kind of doping which stabilizes the two-dimensional boron network~\cite{ismail}. 
During a 
long period the cage structure was not destroyed in the minima hopping run. Instead minima hopping explored the defect structures 
that we have described previously~\cite{pochet} as well as other cage structures which are slightly lower in energy than the 
Szwacki fullerene. Since there is a very large number of possible defect structures 
this cage funnel contains a very large number of local minima and it takes long for minima hopping to escape from it.

\begin{figure} 
\includegraphics[width=0.35\textwidth]{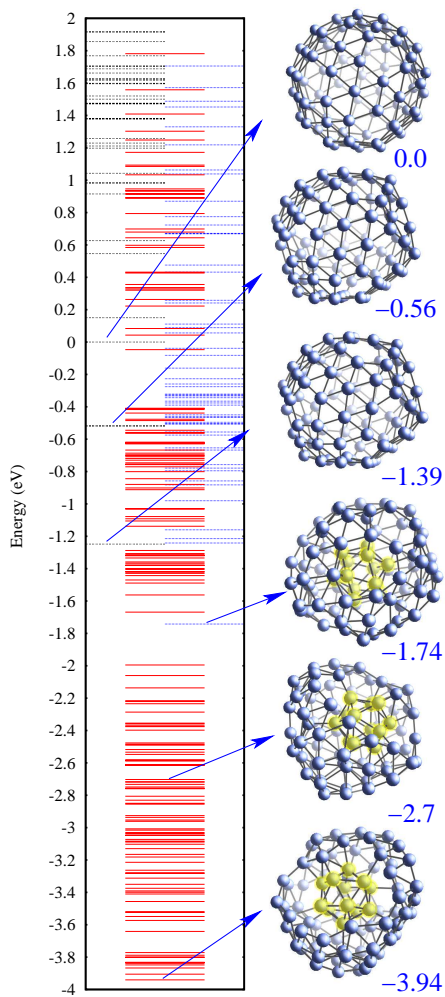}
\caption{\label{B80spec} The configurational energy spectrum of $B_{80}$. 
The energy of the Szwacki fullerene is taken to be zero. The energy levels of the icosahedron-dome 
structures are centered whereas the levels shifted to the left are fullerene like structures.The levels on the right correspond 
to centered icosahedron structures. The atoms of the icosahedra are shown in yellow.
The structure at an energy of -2.7 eV is the putative global minimum from ref~\cite{chinese}.
The energy per atom of our lowest energy $B_{80}$ structure is about .13 eV per atom higher in energy 
than the sheet structure of Tang and Ismail-Beigi~\cite{ismail}.}
\end{figure}

Once one escapes from the fullerene funnel one finds significantly lower energy structures. These structures contain the 
icosahedral $B_{12}$ motif which is the basic building block of elemental boron. This icosahedron is in most cases at the base of a dome like 
structure  or otherwise at the center of a spherical cage. Both the domes and the cages consist mainly but not exclusively of filled and empty 
hexagons and pentagons. Fig.~\ref{B80spec} shows the configurational density of states for the $B_{80}$ cluster. 
The majority of the structures are of the dome type 
and the energies of dome type and fullerene type structures overlaps. Like for the medium size boron clusters many structural building 
blocks can be combined to form clusters of very similar energy. Hence the energy difference between the low energy isomers is 
again very small. 
The lowest energy structure we found is considerably  lower in energy than the recently proposed compact $B_{80}$ structure~\cite{chinese}, 
both within the LDA and PBE functionals.

\begin{figure} [!h]
\begin{center}
\includegraphics[width=0.40\textwidth]{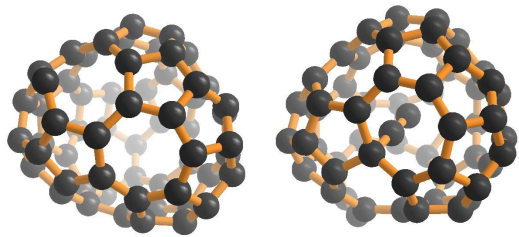}
\caption{\label{C60spec} Two high energy $C_{60}$ cage structures. 
The structure on the left has only 3 fold coordinated atoms even though it contains two 7-member rings.  It is 20.5 eV above the ground state. 
Structures that are even higher in energy can possess some chains with 2-fold coordination and anchor atoms for these chains 
with 4-fold coordination. The structure on the right is an example of such a cage and is 25 eV higher than the ground state. }
\end{center}
\end{figure}

Let us contrast the configurational energy spectrum of $B_{80}$ clusters with the one of $C_{60}$ clusters. 
For $C_{60}$ the first meta-stable structure is a Stone-Wales~\cite{stone} point defect  which 
is 1.6 eV higher in energy than the fullerene ground state. Various defects can be combined to form cages of higher and higher 
energy. Two high energy  structures are shown in Fig.~\ref{C60spec}. The lowest non-cage like structures are however some 25 eV higher in energy than the ground state. This shows that 
in contrast to $B_{80}$ the cage like and non-cage like structures are widely separated in energy. There is consequently a strong driving 
force towards cage like structures and finding the ground state for $C_{60}$ is much easier than for $B_{80}$. 

The differences in the potential energy landscape between $B_{80}$ and  $C_{60}$ are also well illustrated by the following computer experiment. 
If one does a local geometry optimization for 80 boron atoms  starting from random positions one obtains disordered structures 
which are already fairly low in energy, 
namely about 10 eV higher than the ground state. This is in contrast to the case of 60 carbon atoms where a local geometry 
optimization starting from random 
positions gives structures which are about 50 eV above the ground state unless they happen to be cage like. This shows again that the 
boron potential energy landscape has a glassy character with a lot of disordered low energy structures. The energy landscape of 
$C_{60}$ on the other hand has a broad and deep funnel which leads to the ground state fullerene.

\begin{figure} 
\includegraphics[width=0.35\textwidth]{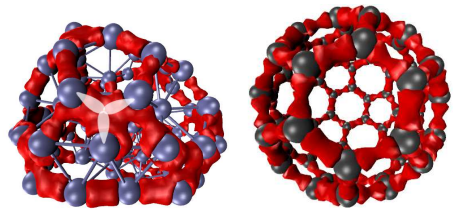}
\caption{\label{rho} The valence charge density in our lowest $B_{80}$ cluster (left) and the $C_{60}$ fullerene (right). 
Whereas in $C_{60}$ we see only two center bonds, both 2 and 3 center bonds are visible in $B_{80}$.}
\end{figure}

The glassy energy landscape of bulk boron has been explained by the frustrated bonding features of boron where 2-center bonds 
have to coexist with 3-center bonds~\cite{oganov}. The glassy energy landscape of the medium size boron clusters can also be explained in 
this way. Fig.~\ref{rho} shows the coexistence of these two types of bonds in our lowest energy $B_{80}$.

In addition to the $B_{80}$ cluster we also examined the $B_{92}$ and $B_{100}$ cluster. 
A structure with a icosahedron in the center of a 80 atom Szwacki fullerene is 7.7 eV lower than the fullerene which was obtained 
by filling the 12 pentagons~\cite{Szwacki}. The resulting structure has however not anymore a high symmetry. 
A stuffed fullerene structure was  proposed for $B_{100}$~\cite{prasad}. Doing minima hopping runs starting from this 
configuration some structures with lower energy and lower symmetry were found as well. These structures were also about 10 eV 
lower in energy than  the recently proposed $B_{100}$ fullerene \cite{boustani2}.
This shows that disordered cages with an icosahedron inside are the basic structural motif for boron clusters in this size range. 

Among all the ground state structures of boron clusters of any size, we could not find any high symmetries.  
Hence the vibrational modes have no or only low degeneracy.
Following these modes by some mode following techniques will therefore in general lead to different transition states 
with different barrier heights. Since the height of the barrier correlates with curvature along the starting mode~\cite{correlate} , 
one can expect for a cluster of low symmetry a broader distribution of barrier heights and therefore 
a larger probability of finding low energy barriers~\cite{lowbar}. If low barriers exist a small modification of the external environment such as 
the presence of another cluster can make these barriers disappear. Hence it is not surprising that all boron structures that we examined, 
independently of whether they are medium size, large, cage-like or not, turned out to be chemically reactive with other boron clusters when 
they are brought into contact. During such a chemical reaction with another cluster several chemical bonds are formed which 
leads to a considerable lowering of the energy and to a large distortion or even destruction of the original structures. This means that even though 
medium size clusters have a strong tendency for cage formation in isolation, it is unlikely that such boron cages exist in nature.
This behavior is also in contrast to the behavior of the $C_{60}$ and $B_{16}N_{16}$
fullerenes. They are only weakly interacting and do not form chemical bonds when they are brought into contact. 
The chemical reactivity of the boron clusters can also be rationalized in a local picture. If many different structural motifs can be used as 
a building block of a low symmetry cluster, it is very likely that some atoms have some dangling bonds which are chemically reactive. 

Our results explain why boron fullerenes have not been found experimentally. Boron clusters are frustrated systems which do not have enough 
electrons to fill all electronic orbitals in a chemical bonding based on pure sp2 hybridization and they consequently do 
not exhibit some clear preference for a simple structural motif. Hence, from a energetical perspective, there is
no driving force towards some well defined structure.
Instead one finds a glassy energy landscape with a large number of different low energy structures whose energies are very similar. 
These structures are chemically reactive and will therefore not be found under experimental conditions. 
The fact that no elemental boron but only compounds containing boron can be found on earth 
however indicates the possibility of synthesizing more complicated boron cages such
as metal doped boron fullerenes. 
Such a doping can energetically  pull down  the cage like part of the configurational space of boron clusters~\cite{pochet}. 

Our simulations demonstrate that one can make theoretical predictions about the feasibility of an experimental synthesis. 
In order to judge whether a system can be formed in nature , it is not necessary to simulate its synthesis process explicitly 
by molecular dynamics or similar methods.
A global geometry optimization with the Minima Hopping algorithm indicates whether the system being simulated 
is a structure seeker or a system with a  glass like potential energy surface. For a glassy system finding the global minimum is slow 
because one has to explore energetic regions with a large density of minima whose energies are very similar.
For a structure seeker on the other hand the energy goes down rapidly and by significant amounts as one approaches the ground state. 
Only for these latter systems it is to be expected that synthesis pathways can be found.

Our work thus clearly shows that theoretical cluster structure prediction has to be based on global geometry optimization 
because only this approach gives the necessary information on the potential energy landscape. 
The standard approach based on structures obtained from educated guesses, that were subsequently locally relaxed, gives only a very incomplete 
characterization of a system. A ground state structure predicted by global geometry optimization has a reasonable chance of being found 
in nature in significant quantities only if it is a) at the bottom of a broad and deep funnel, b) is significantly lower in energy than 
the other low energy meta-stable structures and c) has high symmetry. 

We thank the Indo-Swiss Research grant and SNF for the financial support. 
This work was performed using HPC resources from CSCS and GENCI-CINES (Grant 2010-c2010096194).


\begin{thebibliography}{99}

\bibitem{predict} 
Kagaku, Vol. 25, p.854 (1970) ; 
 Thrower, P. A. (1999). "Editorial". Carbon 37: 1677–1678

\bibitem{c60discovery} 
 Kroto et al. (1985). . Nature 318: 162–163.

\bibitem{list} 
 E. Spano et al., J. Phys. Chem. B 2003, 107, 10337;  
 B. Wang et al.,  J. Phys. Chem. C 2010, 114, 5741 ; 
 T. Oku et al.,  Chemical Physics Letters 380 (2003) 620; 
 Phys. Rev. Lett. 87, 045503 (2001) V. Kumar et al.
 Phys. Rev. Lett. 88, 235504 (2002) V. Kumar et al.
 Phys. Rev. Lett. 90, 055502 (2003) V. Kumar et al. 
 Nano Lett. 4, 677 (2004). V. Kumar et al.
 Phys. Rev. B72, 125427 (2005) Young-Cho Bae et al 
 Phys. Rev. B79, 085423 (2009), V. Kumar 
J. U. Reveles et al. PNAS  December 5, 2006   vol. 103  no. 49  18405-18410
J. Ulises Reveles et al. ,J. Phys. Chem. C 2010, 114, 10739–10744
Satya Bulusu et al. PNAS  May 30, 2006   vol. 103  no. 22  8326-8330
Johansson M. P. et al. (2004) Angew. Chem. Int. Ed 43:2678–2681
M. Walter et al. , Phys. Chem. Chem. Phys. 2006
P. Pochet et al., Rev. B 82, 035431 (2010). 


\bibitem{si20}
A. Willand et al, Phys. Rev. B {\bf 81}, 201405(R) (2010)


\bibitem{Szwacki}
Szwacki et al., Phys. Rev. Lett. 98, 166804 (2007)


\bibitem{Bfull}
R. Zope, Europ. Phys. Lett. 85 68005-p1 (2009)  ; 
N. G. Szwacki, Nanoscale Res. Lett. 3, 49
(2008).

\bibitem{sBfull}
D. L. V. K. Prasad and E. D. Jemmis, Phys. Rev. Lett. 100, 165504 (2008)

\bibitem{chinese}
Jijun Zhao, Lu Wang†, Fengyu Li, and Zhongfang Chen
J. Phys. Chem. A, 2010, 114 (37), pp 9969–9972

\bibitem{minhop}
S. Goedecker, J. Chem. Phys. {\bf 120}, 9911 (2004);
S. Goedecker, W. Hellmann, and T. Lenosky, Phys. Rev. Lett. {\bf 95}, 055501 (2005).

\bibitem{bigdft}
L. Genovese {\em et al.}, J. Chem. Phys. {\bf 129}, 014109 (2008).
 
 
 \bibitem{gth}
S. Goedecker, M. Teter, and J. Hutter, Phys. Rev. B {\bf 54}, 1703 (1996).

\bibitem{pbe}
J. Perdew, K. Burke, and M. Ernzerhof, Phys. Rev. Lett. {\bf 77}, 3865 (1996).


\bibitem{goldberg}
Zhi Xu, Dmitri Golberg, Yoshio Bando
Chemical Physics Letters 480, 110  (2009) 
 
\bibitem{trend}
Yafei Li et al., J. of Chem. Phys., 130, 204706 (2009)

 \bibitem{medium}
L. Wang, J, Zhao, F. li and Z. Chen, Chem. Phys. Lett. 501, 16 (2010)

\bibitem{boustani}
I. Boustani, Phys. Rev. B 55, 16426 (1997)

\bibitem{ismail}
H. Tang and S. Ismail-Beigi, Phys. Rev. Lett. 99, 115501
(2007)




\bibitem{pochet}
Pochet et al., Phys. Rev. B 83, 081403(R) (2011)

\bibitem{stone}
A. Stone and D. Wales, Chem. Phys. Lett. {\bf 128} 501 (1986)

\bibitem{oganov} A. Oganov et al., Nature 457 863 (2009)
\bibitem{prasad}
Prasad et al. ,Phys. Rev. Lett. 100,165504 (2008)
\bibitem{boustani2}
C. \"Ozdo\ifmmode \breve{g}\else \u{g}\fi{}an, S. Mukhopadhyay, M. Hayami, Z. G\"uven\ifmmode \mbox{\c{c}}\else \c{c}\fi{}, R. Pandey, and I. Boustani  , J. Phys. Chem. C 114, 4362 (2010)


\bibitem{correlate}
D. Wales, Science 293 page 2067 2001



\bibitem{lowbar}
A. Heuer, Journal of Physics: Condensed Matter 20, 373101 (2008) ; 
G. Daldoss, O. Pilla, G. Viliani, C. Brangian and G. Ruocco, Phys. Rev. B {\bf 60} 3200 (1999)




















\end{thebibliography}
\end{document}